\documentclass[10pt,twocolumn,conference]{IEEEtran}
\usepackage{times}
\usepackage{amsbsy}
\usepackage{amssymb}
\usepackage{fontenc}
\usepackage[usenames]{color}
\usepackage{latexsym}
\usepackage{amsmath}
\usepackage[ansinew]{inputenc}
\usepackage[dvips ]{graphicx}
\usepackage{float}
\input epsf
\usepackage{epic,eepic}
\usepackage{url}

\IEEEoverridecommandlockouts

\title{\huge Low-Complexity Lattice Reduction-Aided Channel Inversion Methods for Large-Dimensional Multi-User MIMO Systems \vspace{-0.5em}}
\author{\textit{Keke Zu}$^\text{\textdagger}$, \textit{Rodrigo C.\ de Lamare}$^\text{\textdagger}$ \textit{and} \textit{Martin Haardt}$^\text{\textdaggerdbl}$\\
\textdagger~ Communications Research Group, Department of Electronics, University of York\\
York Y010 5DD, United Kingdom\\
\textdaggerdbl~ Communications Research Laboratory, Ilmenau University of Technology \\
PO Box 100565, D-98684 Ilmenau, Germany \\
Emails: \{kz511, rcdl500\}@ohm.york.ac.uk,
martin.haardt@tu-ilmenau.de}

\begin{document}
\maketitle \thispagestyle{empty} \vspace*{-1.75em}

\begin{abstract}
Low-complexity precoding { algorithms} are proposed in this work to
reduce the computational complexity and improve the performance of
regularized block diagonalization (RBD) { based} precoding {
schemes} for large multi-user { MIMO} (MU-MIMO) systems. The
proposed algorithms are based on a channel inversion technique, QR
decompositions{ ,} and lattice reductions to decouple the MU-MIMO
channel into equivalent SU-MIMO channels. Simulation results show
that the proposed precoding algorithms can achieve almost the same
sum-rate performance as RBD precoding, substantial bit error rate
(BER) performance gains{ ,} and a simplified receiver structure,
while requiring a lower complexity.
\end{abstract}

\section{Introduction}
Block diagonalization (BD) { based precoding} techniques
\cite{Spencer,Choi} { are} well-known precoding strateg{ ies} for
multi-user multiple-input multiple-output (MU-MIMO) systems. By
implementing two SVD operations, BD precoding can eliminate the
multi-user interference (MUI). Since BD precoding { focuses} on
canceling { the} MUI, it suffers a performance loss at low signal to
noise ratios (SNRs) when the noise is the dominant factor. By
relaxing the zero MUI constraint, the regularized block
diagonalization (RBD) precoding { scheme} has been proposed in
\cite{RBD}. Instead of achieving strictly independent parallel
channels between { the} users as BD precoding, RBD precoding allows
a small level of interference between { the} users. Although a
better performance is obtained by { the} RBD precoding, it still
needs two SVD operations as BD precoding. As revealed in this paper,
the computational complexity of the RBD precoding algorithm depends
on the number of users and { the} dimensions of each user's channel
matrix which could result in a considerable computational cost for
large MIMO systems. The high cost of the two SVD operations required
by the RBD precoding suggests that precoding algorithms { with}
lower complexity should be investigated for use in { very} large
MIMO systems.

In order to reduce the computational complexity, a generalized MMSE
channel inversion (GMI) precoding algorithm has been proposed in
\cite{GMI} to implement the RBD precoding { scheme}. The first SVD
operation of the RBD precoding is implemented by a matrix inversion
method in GMI precoding. In \cite{QRRBD}, the first SVD operation of
the RBD precoding is replaced with a less complex QR decomposition,
and we term it as QR/SVD RBD precoding. For the second SVD
operation, however, both { the} GMI and { the} QR/SVD RBD precoding
algorithms require the same number of operations as the { original}
RBD precoding { scheme}. If the second SVD operation is implemented
at the transmit side, then the corresponding unitary decoding matrix
{ needs} to be known by each distributed receiver, which requires an
extra control overhead \cite{Chae}. In this work, we develop a
simplified GMI (S-GMI) method to obtain the first precoding filters.
In order to reduce the complexity further and to obtain a better BER
performance, we transform the equivalent SU-MIMO channels into the
lattice space { after the first precoding process} by utilizing the
lattice reduction (LR) technique \cite{CLLL} whose complexity is
mainly due to a QR decomposition. Then, a linear precoding algorithm
is employed instead of the second SVD operation to parallelize each
user's streams.

The essential premise of using transmit processing techniques is the
knowledge of the channel state information (CSI) at the transmitter
{ \cite{Spencer} - \cite{Chae}}. In time-division duplexing (TDD)
systems, CSI can be obtained at the BS by exploiting reciprocity
between the forward and reverse links. In frequency-division
duplexing (FDD) systems, reciprocity is usually not available, but
the BS can obtain knowledge of the downlink user channels by
allowing the users to send a small number of feedback bits on the
uplink { \cite{Rodrigo, Cai}}. We assume that { full} CSI is
available at the transmit side since limited feedback technique{ s
are} not the main focus of this work. In this context, it is worth
noting that the two SVD operations and the decoding matrix at each
receiver are no longer required. The computational complexity is
reduced and the receiver structure can be simplified. A significant
amount of power can be saved which is very important considering the
mobility of distributed users. For convenience, the proposed
precoding algorithm is abbreviated as LR-S-GMI. According to the
specific precoding constraint, the proposed LR-S-GMI precoding
algorithms are categorized as LR-S-GMI-ZF and LR-S-GMI-MMSE
precoding, respectively. We compare the proposed LR-S-GMI {
technique} to the precoding algorithms reported in the literature
including the BD, RBD, QR/SVD RBD and GMI precoding algorithms.

This paper is organized as follows. The system model is given in
Section II. A brief review of the RBD precoding algorithms is
presented in Section III. The proposed LR-S-GMI precoding algorithms
are described in detail in { Section~IV}. Simulation results and
conclusions are presented in { Section~V} and Section VI,
respectively.


\section{System Model}
  We consider an uncoded MU-MIMO downlink channel, with $N_T$ transmit antennas
at the base station (BS) and $N_i$ receive antennas at the $i$th
user equipment (UE). With $K$ users in the system, the total number
of receive antennas is $N_R=\sum _{i=1}^{K}N_i$. A block diagram of such a system is illustrated in Fig. 1.
\begin{figure}[htp]
\begin{center}
\def\epsfsize#1#2{1.0\columnwidth}
\epsfbox{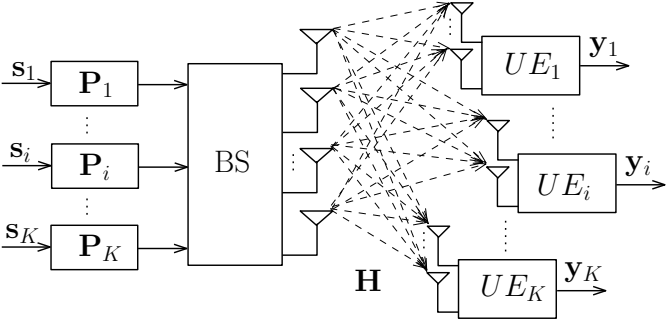} \vspace{-0.8em} \caption{MU-MIMO System Model}
\end{center}
\end{figure}
From the system model, the combined channel matrix $\boldsymbol H$
and the { combined} precoding matrix $\boldsymbol P$ of all users
are given by
 \begin{align}
&\boldsymbol H=[\boldsymbol H_1^T ~ \boldsymbol H_2^T ~ \ldots ~
\boldsymbol H_K^T]^T\in\mathbb{C}^{N_R\times N_T},\\
&\boldsymbol P=[\boldsymbol P_1 ~ \boldsymbol P_2 ~ \ldots ~
\boldsymbol P_K]\in\mathbb{C}^{N_T\times N_R},
\end{align}
where ${\boldsymbol H_i}\in\mathbb{C}^{N_i\times N_T}$ is the $i$th user's channel matrix.
The quantity ${\boldsymbol P_i}\in\mathbb{C}^{N_T\times N_i}$ is the $i$th user's precoding
matrix. We assume a flat
fading MIMO channel and the received signal $\boldsymbol y_i\in\mathbb{C}^{N_i}$ at the $i$th user is given by
  \begin{align}
  \boldsymbol y_i =\boldsymbol H_i\boldsymbol x_i+\boldsymbol H_i\sum _{j=1,j\neq i}^{K}\boldsymbol x_j + {\boldsymbol n}_i,
  \end{align}
  where the quantity ${\boldsymbol x_i}\in\mathbb{C}^{N_i}$ is the
$i$th user's transmit signal, and ${\boldsymbol
n}_i\in\mathbb{C}^{N_i} $ is the $i$th user's Gaussian noise with
independent and identically distributed (i.i.d.) entries of zero
mean and variance $\sigma_n^2$. Assuming that the average transmit
power for { user $i$} is $E_{s_i}$, we construct a normalized signal
$\boldsymbol x_i$ such that
\begin{align}
{ \boldsymbol x_i={\boldsymbol s_i \over \sqrt {{\gamma_i}}}},
\end{align}
where $\boldsymbol s_i=\boldsymbol P_i\boldsymbol d_i$ with
$\boldsymbol d_i$ { being} the data vector{ ,}
$\gamma_i=\|\boldsymbol s_i\|_2^2/E_{s_i}$. { With this
normalization, the transmit signal $\boldsymbol x_i$ obeys ${\rm
E}\| \boldsymbol x_i\|^2_2=E_{s_i}$ \cite{Hochwald}.}

The received signal $\boldsymbol y_i$ is weighted by the scalar $\sqrt \gamma_i$ to form the estimate
  \begin{align}
  \boldsymbol {\hat d}_i =\sqrt \gamma_i\boldsymbol y_i,
  \end{align}
where the physical meaning of the scalar $\sqrt\gamma_i$ is to make
sure { that} the average transmit power $E_{s_i}$ is still the same
after the precoding process. Note that it is necessary to cancel
$\sqrt\gamma_i$ out at the receiver to get the correct amplitude of
the desired signal part.

\section{Review of RBD Precoding Algorithm}
The design of { the} RBD precoding algorithm is performed in two
steps \cite{RBD}. The first precoding filter is used to balance the
MUI with noise, then { approximate} parallel SU-MIMO channels are
obtained. The second precoding filter is implemented to parallelize
each user's streams. Correspondingly, the precoding matrix
$\boldsymbol P$ can be rewritten as
\begin{align}
\boldsymbol P= {\boldsymbol P}^a{\boldsymbol P}^b,
\end{align}
where ${\boldsymbol P}^a=[{\boldsymbol P}_1^a~{\boldsymbol P}_2^a\ldots{\boldsymbol P}_K^a]$ and $\boldsymbol  {P}^b={\rm diag}\{\boldsymbol{P}_1^b,\boldsymbol{P}_2^b,\ldots,\boldsymbol{P}_K^b$\}.
%
We exclude the $i$th user's channel matrix
and define $\boldsymbol{\overline H}_i$ as
\begin{align}
\boldsymbol{\overline H}_i=[\boldsymbol H_1^T ~\dots~\boldsymbol H_{i-1}^T ~ \boldsymbol H_{i+1}^T ~ \ldots ~ \boldsymbol
H_K^T]^T\in\mathbb{C}^{\overline N_i\times N_T},
\end{align}
where $\overline N_i=N_R-N_i$. Thus, the interference generated to the other users is determined by $\boldsymbol{\overline H}_i\boldsymbol P_i^a$.
In order to balance the MUI and the noise term, an RBD constraint is developed in \cite{RBD} and given by
\begin{align}
\boldsymbol P_i^a = \min_{\boldsymbol P_i^a} \rm E\lbrace\sum_{i=1}^K\|\boldsymbol{\overline H}_i\boldsymbol P_i^a\|^2+{\gamma\|\boldsymbol n\|_F^2}\rbrace\nonumber\\
{\rm s.t.}~\rm E\|\boldsymbol x\|_F^2=E_s.
\end{align}
{ Assuming} that the rank of $\boldsymbol{\overline H}_i$ is
$\overline L_i$, define the SVD of $\boldsymbol{\overline H}_i$
\begin{align}
\boldsymbol{\overline H}_i=\boldsymbol{\overline U}_i\boldsymbol{\overline \Sigma}_i\boldsymbol{\overline V}_i^H=\boldsymbol{\overline U}_i\boldsymbol{\overline \Sigma}_i[\boldsymbol{\overline V}_i^{(1)}~\boldsymbol{\overline V}_i^{(0)}]^H,
\end{align}
where $\boldsymbol{\overline U}_i\in\mathbb{C}^{\overline N_i\times \overline N_i}$ and $\boldsymbol{\overline V}_i\in\mathbb{C}^{N_T\times N_T}$ are unitary matrices. The diagonal matrix $\boldsymbol{\overline \Sigma}_i \in\mathbb{C}^{\overline N_i\times N_T}$ contains the singular
values of the matrix $\boldsymbol{\overline H}_i$.
As shown in \cite{RBD}, the solution for the RBD constraint can be obtained as
\begin{align}
{\boldsymbol P_i^a}^{\rm (RBD)}=\boldsymbol{\overline V}_i(\boldsymbol{\overline \Sigma}_i^T\boldsymbol{\overline \Sigma}_i+\alpha\boldsymbol I_{N_T})^{-1/2},
\end{align}
where $\alpha={N_R\sigma_n^2\over E_s}$ is the regularization parameter.

After the first RBD precoding process, the MU-MIMO channel is
decoupled into a set of $K$ { approximately} parallel SU-MIMO
channels. Due to the regularization process, there are small
residual interferences between these channels, and these
interferences tend to zero at high SNRs. Thus, the effective channel
matrix for the $i$th user can be expressed as
\begin{align}
\boldsymbol H_{{\rm eff}_i}=\boldsymbol H_i\boldsymbol P_i^a.
\end{align}
Define $L_{\rm eff}={\rm rank}(\boldsymbol H_{{\rm eff}_i})$ and consider the second SVD operation on the effective channel matrix
\begin{align}
\boldsymbol H_{{\rm eff}_i}=\boldsymbol{U}_i\boldsymbol{\Sigma}_i{\boldsymbol{V}_i}^H,
\end{align}
{ using} the unitary matrices $\boldsymbol{U}_i
\in\mathbb{C}^{L_{\rm eff}\times L_{\rm eff}}$ and
$\boldsymbol{V}_i\in\mathbb{C}^{N_T\times N_T}$. The second
precoding filters for RBD precoding can be obtained as
\begin {align}
{\boldsymbol P_i^b}^{\rm (RBD)}=\boldsymbol{V}_i{\boldsymbol\Lambda}^{\rm (RBD)},
\end {align}
where ${\boldsymbol\Lambda}$ is the power loading matrix that
depends on the optimization { criterion}. An example power loading
is the water filling (WF) \cite{Paulraj01}. The $i$th user's
decoding matrix is obtained as
\begin {align}
{\boldsymbol G_i}=\boldsymbol{U}_i^H,
\end {align}
which needs to be known by each user's receiver.

Note that for the conventional RBD precoding algorithm, there is a {
dimensionality} constraint to be satisfied
\begin {align}
N_T>{\rm max}\{{\rm rank}(\boldsymbol{\overline H}_1),{\rm rank}(\boldsymbol{\overline H}_2),\ldots,{\rm rank}(\boldsymbol{\overline H}_K)\}.
\end {align}
Then, we can get the matrix dimension relationship as $L_{\rm
eff}\leq N_i<\overline N_i<N_R\leq N_T$. Note that the first SVD
operation in (9) needs to be implemented $K$ times on
$\boldsymbol{\overline H}_i$ with dimension { $\overline N_i\times
N_T$} and the second SVD operation in (12) needs to be implemented
$K$ times on $\boldsymbol H_{{\rm eff}_i}$ with dimension { $L_{\rm
eff}\times N_T$}. From the above analysis, most of the computational
complexity of the RBD precoding algorithm comes from the two SVD
operations which make the computational complexity of the RBD
precoding algorithm { increase with} the number of users $K$ and the
system dimensions. In order to reduce the computational complexity
of the RBD precoding algorithm, low complexity precoding algorithms
for MU-MIMO systems are proposed in what follows.

\section{Proposed Low Complexity LR-S-GMI Precoding Algorithms}
In this section, we describe the proposed low-complexity LR-S-GMI precoding algorithms based on a strategy that employs a
channel inversion method \cite{GMI}, QR decompositions, and lattice reductions. Similar to the RBD precoding algorithm, the design of the proposed LR-S-GMI precoding algorithms is computed in two steps.

First, we obtain $\boldsymbol P_i^a$ in the conventional RBD precoding algorithm for the LR-S-GMI precoding algorithms by using one channel inversion and $K$ QR decompositions.
By applying the MMSE channel inversion, we have
\begin{equation}
\begin{split}
\boldsymbol H^{\dag}_{{\rm mse}} & =(\boldsymbol H^H\boldsymbol
H+\alpha \boldsymbol I)^{-1}\boldsymbol H^H \\ & =[{ \boldsymbol
H_{1,{\rm mse}},\boldsymbol H_{2,{\rm mse}},\ldots,\boldsymbol
H_{K,{\rm mse}}}].
\end{split}
\end{equation}
Considering a high SNR case, it can be shown that ${\boldsymbol H}
{\boldsymbol H}^\dag_{{\rm mse}} \approx \boldsymbol I_{N_T}$
\cite{Michael}. This means that the off-diagonal block matrices of
${\boldsymbol H} {\boldsymbol H}^\dag_{{\rm mse}}$ converge to zero
as the SNR increases. Then, we obtain a condition which shows { that
${\boldsymbol H}_{i,{\rm mse}}$} is in the null space of
${\boldsymbol{\overline H}}_i$
\begin{align}
{\boldsymbol{\overline H}}_i { {\boldsymbol H}_{i,{\rm mse}}}\approx
\boldsymbol 0.
\end {align}
Considering the QR decomposition of { $\boldsymbol H_{i,{\rm
mse}}=\boldsymbol Q_{i,{\rm mse}}\boldsymbol R_{i,{\rm mse}}$}, we
have
\begin{align}
\boldsymbol{\overline H}_i\boldsymbol H_{i,{\rm mse}} =
\boldsymbol{\overline H}_i\boldsymbol Q_{i,{\rm mse}} \boldsymbol
R_{i,{\rm mse}} \approx \boldsymbol 0 ~{\rm for}~i=1,\ldots,K,
\end{align} where $\boldsymbol R_{i,{\rm
mse}}\in\mathbb{C}^{N_i\times N_i}$ is an upper triangular matrix and
${\boldsymbol Q}_{i,{\rm mse}}\in\mathbb{C}^{N_T\times N_i}$ is an orthogonal
matrix. Since $\boldsymbol R_{i,{\rm mse}}$ is invertible, we
have
\begin{align}
\boldsymbol{\overline H}_i\boldsymbol Q_{i,{\rm
mse}}\approx \boldsymbol 0.
\end {align}
Thus, $\boldsymbol Q_{i,{\rm mse}}$ satisfies the RBD constraint (8)
to balance the MUI and { the} noise.

We have simplified the design of $\boldsymbol P_i^a$ for the RBD precoding here as compared to \cite{GMI} where a
residual interference suppression filter $\boldsymbol T_i$ is
applied after $\boldsymbol P_i^a$. The filter $\boldsymbol T_i$ increases the complexity and
cannot completely cancel the MUI. Therefore, we omit the residual
interference suppression part since it is not necessary for the RBD
precoding. We term the simplified GMI as S-GMI precoding in this work.
Then, the first precoding matrix can be obtained as
\begin{align}
\boldsymbol P_i^a=\boldsymbol Q_{i,{\rm mse}},
\end{align}
and the first combined precoding matrix is
\begin{align}
\boldsymbol P^a=[\boldsymbol P_1^a,~\boldsymbol P_2^a,~\ldots,~\boldsymbol P_K^a].
\end{align}
%

Next, we employ the LR-aided linear precoding algorithm instead of
the second SVD operation to obtain $\boldsymbol P_i^b$ { and
parallelize} each user's streams. The aim of the LR transformation
is to find a new basis $\boldsymbol {\tilde H}$ which is nearly
orthogonal compared to the original matrix $\boldsymbol H$ for a
given lattice $L(\boldsymbol H)$. The most commonly used LR
algorithm has been first proposed by Lenstra, Lenstra and L. {
Lov\'asz} (LLL) in \cite{LLL} with polynomial time complexity. In
order to reduce the computational complexity, a complex LLL (CLLL)
algorithm was proposed in \cite{CLLL}, which reduces the overall
complexity of the LLL algorithm { by nearly half} without
sacrificing any performance. We employ the CLLL algorithm to
implement the LR transformation in this work.

After the first precoding, we transform the MU-MIMO channel into approximate parallel SU-MIMO channels and the effective channel matrix for the $i$th user is
\begin{align}
\boldsymbol H_{{\rm eff}_i}=\boldsymbol H_i\boldsymbol P_i^a.
\end{align}
We perform the LR transformation on ${\boldsymbol H}_{{\rm
eff}_i}^T$ in the precoding scenario \cite{Windpassinger}, that is
\begin{align}
{\boldsymbol {\tilde H}}_{{\rm eff}_i}=\boldsymbol T_i\boldsymbol
H_{{\rm eff}_i}~{\rm and}~ \boldsymbol H_{{\rm eff}_i}=\boldsymbol T_i^{-1}{\boldsymbol {\tilde H}}_{{\rm eff}_i},
\end{align}
where $\boldsymbol T_i$ is a unimodular matrix ($\rm{det}|\boldsymbol
T_i|=1$) and all elements of $\boldsymbol T_i$ are complex
integers, i.e. $ t_{l,k} \in\mathbb{Z}+j\mathbb{Z}$.

Following the LR transformation, we employ the linear precoding constraint to get the second precoding matrix instead of the second SVD operation in (12). The ZF precoding constraint is implemented for user $i$ as
\begin{align}
\boldsymbol {\tilde P}_{{\rm ZF}_i}^b={\boldsymbol {\tilde H}}_{{\rm
eff}_i}^H (\boldsymbol {\tilde H}_{{\rm eff}_i}{\boldsymbol {\tilde
H}}_{{\rm eff}_i}^H)^{-1}.
\end{align}
It is well-known that the performance of MMSE precoding is always better than that of ZF precoding.
We can get the second precoding filter by employing an MMSE precoding constraint.
The MMSE precoding is actually equivalent to ZF precoding with respect to an extended
system model \cite {Wuebben,Li}. The extended channel matrix
$\boldsymbol {\underline H}$ for the MMSE precoding scheme is defined as
\begin{align}
\boldsymbol {\underline H}=\begin {bmatrix}\boldsymbol H,\sqrt\alpha \boldsymbol I_{N_R}\end {bmatrix}.
\end{align}
By introducing the regularization factor $\alpha$, a trade-off between the level of MUI and noise is introduced \cite{Michael}. Then, the MMSE precoding filter is obtained as
\begin{align}
\boldsymbol P_{\rm MMSE}=\boldsymbol A {\boldsymbol {\underline H}}^H(\boldsymbol
{\underline H}{\boldsymbol {\underline H}}^H)^{-1},
\end{align}
where $\boldsymbol A=\begin {bmatrix} \boldsymbol I_{N_T},\boldsymbol 0_{N_T,N_R}\end {bmatrix}$, and the multiplication of $\boldsymbol A$ will not result in transmit power amplification since $\boldsymbol A\boldsymbol A^H=\boldsymbol I_{N_t}$.
From the mathematical expression in (26), the rows of $\boldsymbol {\underline H}$ determine the effective
transmit power amplification of the MMSE precoding. Correspondingly, the LR
transformation for MMSE precoding should be applied to the transpose of the extended
channel matrix ${\boldsymbol {\underline H}}_{{\rm
eff}_i}^T={\begin {bmatrix}\boldsymbol H_{{\rm eff}_i},\sqrt\alpha\boldsymbol
I_{N_i}\end {bmatrix}}^T$ for the MMSE precoding, and the LR transformed
channel matrix ${\boldsymbol {\underline {\tilde H}}}_{{\rm eff}_i}$
is obtained as
\begin {align}
{\boldsymbol {\underline {\tilde H}}}_{{\rm eff}_i}=\boldsymbol {\underline T}_i\boldsymbol
{\underline H}_{{\rm eff}_i}.
\end {align}
Then, the LR-aided MMSE precoding filter is given by
\begin{align}
 {\boldsymbol {\tilde P}}_{{\rm MMSE}_i}^b=\boldsymbol
 A_i {\boldsymbol {\tilde{\underline H}}}_{{\rm eff}_i}^H({\boldsymbol
{\tilde{\underline H}}}_{{\rm eff}_i}{\boldsymbol {\tilde{\underline
H}}}_{{\rm eff}_i}^H)^{-1}.
\end{align}
Finally, the second precoding matrix $\boldsymbol {\tilde P}^b$ for all users is
\begin{align}
\boldsymbol  {\tilde P}^b={\rm diag}\{\boldsymbol {\tilde P}_1^b,\boldsymbol  {\tilde P}_2^b,\ldots,\boldsymbol  {\tilde P}_K^b\}.
\end{align}
The resulting precoding matrix is $\boldsymbol {\tilde P}=
\boldsymbol P^a \boldsymbol {\tilde P}^b$. Since the lattice reduced precoding matrix $\boldsymbol
{\tilde P}^b$ has near orthogonal columns, the required transmit power
will be reduced compared to the BD or RBD precoding algorithms. Thus, a better BER performance than the RBD precoding algorithm can be achieved by the proposed LR-S-GMI precoding algorithms.

The received signal is finally obtained as
\begin{align}
\boldsymbol y=\boldsymbol H\boldsymbol {\tilde P}\boldsymbol d+\sqrt\gamma\boldsymbol n.
\end{align}
The main processing work left for the receiver is to quantize the received signal $\boldsymbol y$ to the nearest data vector and the decoding matrix $\boldsymbol G$ in (14) is not needed anymore.

The proposed precoding algorithms are called LR-S-GMI-ZF and LR-S-GMI-MMSE depending on the choice of the second precoding filter as given in (24) and (28), respectively. We will focus on the LR-S-GMI-MMSE since a better performance is achieved. The implementing steps of the LR-S-GMI-MMSE precoding algorithm are summarized in Table 1. By replacing the steps 8 and 9 in Table I with the formulation in (24), the LR-S-GMI-ZF precoding algorithm can be obtained.
\begin{table}[htp]
\caption{The Proposed LR-S-GMI-MMSE Precoding Algorithm} 
\centering 
\begin{tabular}{l l} 
\hline\hline 
Steps & Operations \\ [0.5ex] 
\hline 
\\
& {\bf Applying the MMSE Channel Inversion}~~~~~~~~~~~~~~~~~~~~~\\
(1)& $\boldsymbol H^{\dag}_{{\rm mse}}  =(\boldsymbol H^H\boldsymbol
H+\alpha \boldsymbol I)^{-1}\boldsymbol H^H$\\
(2)& for i~=~1~:~$ K$\\
(3)&~~~~~$[\boldsymbol Q^\dag_{i,{\rm mse}~ \boldsymbol
R^\dag_{i,{\rm mse}}}]={\rm QR}(\boldsymbol H^\dag_{i,{\rm mse}},~0)$\\
(4)&~~~~~$\boldsymbol P_i^a=\boldsymbol Q^\dag_{i,{\rm mse}}$\\
(5)&~~~~~$\boldsymbol H_{{\rm eff}_i}=\boldsymbol H_i\boldsymbol P_i^a$\\
(6)&~~~~~${\boldsymbol {\underline H}}_{{\rm
eff}_i}={\begin {bmatrix}\boldsymbol H_{{\rm eff}_i} ~ \sqrt\alpha\boldsymbol
I_{N_i}\end {bmatrix}}$\\
(7)&~~~~~$[\boldsymbol {\underline T}_i^T~
\boldsymbol{\underline H}_{{\rm eff}_i}^T]={\rm CLLL}({\boldsymbol {\underline {\tilde H}}}_{{\rm eff}_i}^T)$\\
(8)&~~~~~$\boldsymbol A_i=[\boldsymbol I_i~\boldsymbol 0_i]$\\
(9)&~~~~~$ {\boldsymbol {\tilde P}}_{{\rm MMSE}_i}^b=\boldsymbol
 A_i {\boldsymbol {\tilde{\underline H}}}_{{\rm eff}_i}^H({\boldsymbol
{\tilde{\underline H}}}_{{\rm eff}_i}{\boldsymbol {\tilde{\underline
H}}}_{{\rm eff}_i}^H)^{-1}$\\
(10)&end\\
& {\bf Compute the overall precoding matrix}\\
(11)&$\boldsymbol P^a=[\boldsymbol P_1^a,~\boldsymbol P_2^a,~\ldots,~\boldsymbol P_K^a]$\\
(12)&$\boldsymbol  {\tilde P}^b={\rm diag}\{\boldsymbol {\tilde P}_1^b,\boldsymbol  {\tilde P}_2^b,\ldots,\boldsymbol  {\tilde P}_K^b\}$\\
(13)&$\boldsymbol  {\tilde P}=\boldsymbol P^a\boldsymbol  {\tilde P}^b$\\
& {\bf Calculate the scaling factor $\gamma$}\\
(14)&$\gamma=(\|\boldsymbol  {\tilde P}\boldsymbol d\|_F^2/E_s)$\\
& {\bf Get the received signal}\\
(15)& $\boldsymbol y=\boldsymbol H\boldsymbol {\tilde P}\boldsymbol d+\sqrt\gamma\boldsymbol n$\\
& {\bf Transform back from lattice space}\\
(16)&$\boldsymbol {\hat d}=\boldsymbol{\underline T}\lceil \boldsymbol y \rfloor$\\
\\ [1ex] 
\hline 
\end{tabular}
\end{table}

\section{Simulation Results}
A system with $N_T=8$ transmit antennas and $K=4$ users each
equipped with $N_i=2$ receive antennas is considered; this scenario
is denoted as { the} $(2,2,2,2)\times 8$ case. The vector
$\boldsymbol d_i$ of the $i$th user represents data transmitted with
QPSK modulation.

The channel matrix $\boldsymbol H_i$ of the $i$th user is modeled as a complex Gaussian channel matrix with zero mean and unit variance. We assume an uncorrelated block fading channel. We also
assume that the channel estimation is perfect at the receive side
and the feedback channel is error free. The number of simulation
trials is $10^6$ and the packet length is $10^2$ symbols.
The $E_b/N_0$ is defined as $E_b/N_0={N_RE_s\over N_TMN_0}$ with $M$ being the number of transmitted information bits per channel symbol.

Fig. 2. shows the BER performance of the proposed and existing
precoding algorithms. The QR/SVD RBD and GMI precoding algorithms
achieve almost the same BER performance as the conventional RBD
precoding. It is clear that the S-GMI precoding has a better BER
performance compared to BD, RBD, QR/SVD RBD and GMI precoding
algorithms. The proposed LR-S-GMI-MMSE precoding algorithm shows the
best BER performance. At the BER of $10^{-2}$, the LR-S-GMI-MMSE {
precoding} has a gain of more than 5.5 dB compared to the RBD
precoding. It is worth noting that the BER performance of the RBD
precoding is outperformed by the proposed LR-S-GMI-MMSE precoding in
the whole $E_b/N_0$ range and the BER gains become more significant
with the increase of $E_b/N_0$. { The reason why the proposed
LR-S-GMI-MMSE precoding algorithm provides a better BER performance
than the exiting algorithm is because it provides a better channel
quality as measured by the condition number of the effective
channel}.
\begin{figure}[htp]
\begin{center}
\def\epsfsize#1#2{1.0\columnwidth}
\epsfbox{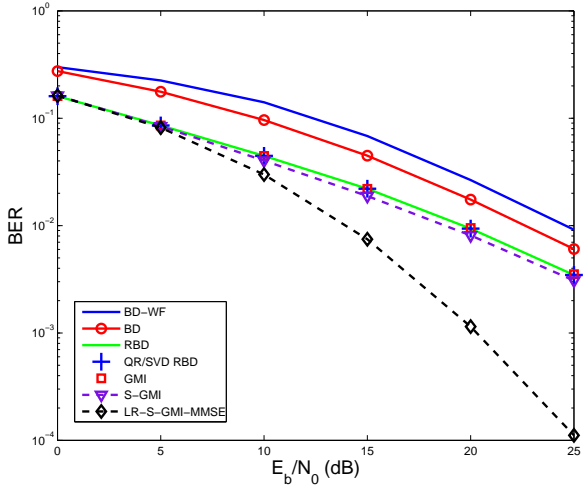} \vspace{-0.8em} \caption{BER performance,
$(2,2,2,2)\times 8$ MU-MIMO}
\end{center}
\end{figure}

Fig. 3. illustrates the sum-rate of the above precoding
algorithms. The information rate is calculated using \cite{Vishwanath}:
\begin{align}
 C={\rm log}({\rm det}(\boldsymbol I+\sigma_n^{-2}\boldsymbol H\boldsymbol P \boldsymbol P^H \boldsymbol H^H)) ~~{\rm (bits/Hz)}.
\end{align}
BD precoding with WF power loading shows a better sum-rate
performance than BD precoding without power loading. However, as
shown in Fig. 2., the BER performance is degraded by applying this
WF scheme. Similar to the BER figure, the RBD, QR/SVD RBD and GMI
precoding algorithms show { a comparable} sum-rate performance. The
S-GMI precoding also achieves the sum-rate performance of the RBD
precoding. The proposed LR-S-GMI-MMSE precoding algorithm shows
almost the same sum-rate performance as the RBD precoding at low
$E_b/N_0$s. At high $E_b/N_0$s, however, the sum-rate performance of
LR-S-GMI-MMSE precoding is slightly inferior to that of the RBD
precoding and approach{ es the performance of} BD precoding.

\begin{figure}[htp]
\begin{center}
\def\epsfsize#1#2{1.0\columnwidth}
\epsfbox{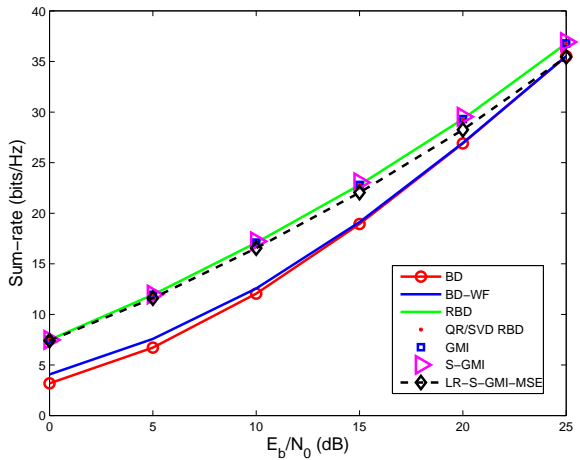} \vspace{-0.8em} \caption{Sum-rate performance,
$(2,2,2,2)\times 8$ MU-MIMO}
\end{center}
\end{figure}

The required floating point operations (FLOPs) for the conventional
BD, RBD and QR/SVD RBD precoding algorithms { are} given in
\cite{Zu, WCNC}. The reduction in the number of FLOPs obtained by
the proposed LR-S-GMI-MMSE is $73.6\%$, $69.5\%$ and $59.1\%$ as
compared to the RBD, BD and QR/SVD RBD precoding algorithms,
respectively.
\section{conclusion}
In this paper, low-complexity precoding algorithms based on a
channel inversion technique, QR decompositions{ ,} and lattice
reductions have been proposed for MU-MIMO systems. The complexity of
the precoding process is reduced and a considerable BER gain is
achieved by the proposed LR-S-GMI precoding algorithms at a cost of
a slight sum-rate loss at high SNRs. Since the proposed LR-S-GMI
precoding algorithms are only implemented at the transmit side, the
decoding matrix is not needed anymore at the receive side compared
to the RBD precoding algorithm. Then, the structure of the receiver
can be simplified, which is an additional benefit of the proposed
LR-S-GMI precoding algorithms.

\end{document}